\documentclass[10pt]{iopart} 
\usepackage{iopams}
\usepackage{amsthm}
\usepackage{bbm}
\usepackage{psfrag}
\usepackage[dvips]{graphicx}
 
\newcommand\NN{{\mathbbm{N}}}
\newcommand\ZZ{{\mathbbm{Z}}}
\newcommand\RR{{\mathbbm{R}}}
\newcommand\dd{{\mathrm{d}}}
\newcommand\ii{{\mathrm{i}}}
\newcommand\ee{{\mathrm{e}}}

\newcommand{\Dirac}{\delta}

\begin{document} 

\title{Microcanonical entropy of the spherical model with nearest-neighbour interactions} 

\author{Michael Kastner$^{1,2}$}
\address{$^1$ National Institute for Theoretical Physics (NITheP), Stellenbosch 7600, South Africa}
\address{$^2$ Institute of Theoretical Physics,  University of Stellenbosch, Stellenbosch 7600, South Africa}
\ead{kastner@sun.ac.za} 

\date{\today}
 
\begin{abstract}
For the spherical model with nearest-neighbour interactions, the microcanonical entropy $s(\varepsilon,m)$ is computed analytically in the thermodynamic limit for all accessible values of the energy $\varepsilon$ and the magnetization $m$ per spin. The entropy function is found to be concave (albeit not strictly concave), implying that the microcanonical and the canonical ensembles are equivalent, despite the long-range nature of the spherical constraint the spins have to obey. Two transition lines are identified in the $(\varepsilon,m)$-plane, separating a paramagnetic phase from a ferromagnetic and an antiferromagnetic one. The resulting microcanonical phase diagram is compared to the more familiar canonical one.
\end{abstract}

\noindent{\it Keywords\/}:  solvable lattice models, classical phase transitions (theory), phase diagrams (theory)

\section{Introduction}
Ensemble averages are at the very core of equilibrium statistical mechanics as introduced by Boltzmann and Gibbs more than a century ago. The crucial idea is that long-time averages, which are the quantities of interest in equilibrium statistical physics, coincide with ensemble averages, with the advantage that the latter are easier to compute. An ensemble average is obtained by assigning statistical weights to all the microstates of a system, and calculating expectation values with respect to these weights. The choice of the statistical ensemble, i.\,e.\ the choice of statistical weights, depends on the physical situation that one is interested in: The microcanonical ensemble for example is appropriate for the description of an isolated system at fixed energy, whereas the canonical ensemble describes a system in equilibrium with an infinitely large heat bath of temperature $T$. For a suitable class of short-range interactions, both ensembles are known to give equivalent results in the thermodynamic limit (see \cite{Ruelle} for details). For actual calculations, equivalence of ensembles allows one to use the ensemble which appears to be the most convenient, independently of whether it reflects the physical situation of interest or not. In this spirit, numerical calculations of microcanonical quantities have been used repeatedly to identify first order phase transitions, as the finite-size precursors of these transitions appear to be more pronounced in the microcanonical ensemble \cite{LeeKosterlitz90,Hueller94,KaProHue:00,Junghans_etal06}. Analytic calculations in the microcanonical ensemble, however, turn out to be more demanding than in the canonical ensemble, as Gibbs already remarked in his seminal treatise \cite{Gibbs02}:
\begin{quote}\it
``It is sufficient here to remark that analytically the canonical distribution is much more manageable than the microcanonical.''
\end{quote}
As a consequence, rather few such microcanonical calculations can be found in the literature. 

In the present paper, a microcanonical analysis is reported for the spherical model with nearest-neighbour interactions on a $d$-dimensional hypercubic lattice. For $d\geqslant3$, this model is a toy model of a ferromagnet, showing a temperature-driven continuous phase transition from a ferromagnetic to a paramagnetic phase. Exposed to a magnetic field and at sufficiently low temperature, the model also undergoes a field-driven discontinuous phase transition. Although the nearest-neighbour interactions are clearly of short-range nature, the model is subject to a spherical constraint (from which it derived its name) which is effectively long-range. In contrast to the short-range interacting case, it cannot be taken for granted that microcanonical and canonical calculations yield equivalent results for long-range interacting systems. As a consequence, it was not clear to the author from the outset whether or not to expect equivalence of microcanonical and canonical calculations for the spherical model \cite{KaSchneSchrei:07,KaSchneSchrei:08}. The possibility to observe such nonequivalence renders the spherical model an interesting case for a study in the microcanonical ensemble.

Behringer \cite{Behringer:05} has published a calculation of this model's microcanonical entropy $s(\varepsilon,m)$, considered as a function of the energy $\varepsilon$ and the magnetization $m$ per spin. However, his analysis is valid only in a certain region of the $(\varepsilon,m)$-plane, leaving out the parameter values in the phase-coexistence region where nonequivalence of ensembles might possibly occur. In this region, a saddle point analysis of the asymptotic integral which yields the microcanonical entropy is no longer possible, and it is one of the main results of the present article to extend the calculation of the microcanonical entropy into the coexistence region.

Apart from completing the microcanonical calculation of $s(\varepsilon,m)$ for the spherical model, the results reported in the present article also serve a more general purpose: from the asymptotic analysis of a Laplace-type integral, one observes that the occurrence of a phase transition is related to the breakdown of the above mentioned saddle point analysis. A careful study of this asymptotic integral nicely illustrates the mathematics at work which leads to a nonanalytic behaviour of the entropy, and hence to a phase transition.

Finally, from the calculation of the microcanonical entropy $s(\varepsilon,m)$ for all accessible values of $\varepsilon$ and $m$, we obtain the microcanonical phase diagram of the spherical model, i.\,e.\ the lines in the $(\varepsilon,m)$-plane for which $s$ is not a smooth function and phase transitions occur. This diagram comprises, as expected, a transition line corresponding to the ferromagnetic to paramagnetic transition of the spherical model, but it also contains a second transition line, signalling a transition from a paramagnetic to an antiferromagnetic phase. Omitting many of the details, this phase diagram, and a similar one for the Ising model, have been published previously as a Letter \cite{KaPlei:09}, suggesting that the observed behaviour is typical for short-range ferromagnets in the microcanonical ensemble.


The rest of the article is organized as follows. In \sref{sec:spherical}, the spherical model is introduced, and a calculation of the microcanonical entropy $s(\varepsilon,m)$ of this model is reported in \sref{sec:micent}. The resulting caloric curves and response functions are discussed in \sref{sec:response}. From the results for the microcanonical entropy, the canonical free energy and the canonical phase diagram are recovered by means of a Legendre transform in \sref{sec:legendre}. Finally, a summary and a discussion of the results is given in \sref{sec:conclusions}.

\section{Spherical model with nearest-neighbour interactions}
\label{sec:spherical}

Consider a hypercubic subset
\begin{equation}
{\mathcal L} = \{1,\dots,L\}^d \subset \ZZ^d
\end{equation}
of a $d$-dimensional hypercubic lattice. On each of the $N=L^d$ lattice sites a real degree of freedom $\sigma_i\in\RR$ is placed. The spherical model is characterized by the Hamiltonian
\begin{equation}\label{eq:H_sph}
H_N:\RR^N\to\RR,\qquad \sigma\mapsto -J\sum_{\langle i,j\rangle} \sigma_i \sigma_j,
\end{equation}
where $J>0$ is a coupling constant and the angular brackets denote a summation over pairs of nearest neighbours on the lattice ${\mathcal L}$. This model was introduced by Berlin and Kac \cite{BerKac:52} as an exactly solvable caricature of the Ising model of a ferromagnet. In contrast to the Ising model case, the ``spin variables'' $\sigma_i$ are real numbers: their modulus is not fixed to unity as for the Ising model where $\sigma_i\in\{-1,+1\}$. Instead, the spherical constraint
\begin{equation}
\sum_{i=1}^N\sigma_i^2=N
\end{equation}
allows for fluctuations of the modulus of the spin variables. In the canonical ensemble, this model has been solved in the thermodynamic limit for arbitrary $d$, and a continuous phase transition from a ferromagnetic phase at low temperatures to a paramagnetic phase at high temperatures (or energies) occurs for all $d\geqslant3$ \cite{Joyce}.

\section{Microcanonical entropy}
\label{sec:micent}

For the spherical model with nearest-neighbour coupling, an analytic calculation of the microcanonical entropy as a function of energy $\varepsilon$ and magnetization $m$ has been reported by Behringer in \cite{Behringer:05}, but the analysis does not apply to certain regions in the $(\varepsilon,m)$-plane. Starting point for such a microcanonical calculation is the density of states
\begin{equation}
\Omega_N(\varepsilon,m)=\int_{\RR^N}\dd\sigma\,\Dirac[N\varepsilon-H_N(\sigma)]\,\Dirac[Nm-M_N(\sigma)]\,\Dirac\bigg(N-\sum_{i=1}^N\sigma_i^2\bigg),
\end{equation}
where the function
\begin{equation}
M_N:\RR^N\to\RR,\qquad \sigma\mapsto \sum_{i=1}^N \sigma_i
\end{equation}
yields the magnetization of a microstate. It was shown in \cite{Behringer:05} that, asymptotically for a large number $N$ of lattice sites, the density of states can be written as
\begin{equation}\label{eq:Omega_3int}
\Omega_N(\varepsilon,m) \sim \!\int_{a-\ii\infty}^{a+\ii\infty}\frac{\dd z}{2\pi} \int_{c-\ii\infty}^{c+\ii\infty}\frac{\dd u}{2\pi} \int_{b-\ii\infty}^{b+\ii\infty}\frac{\dd w}{2\pi} \,\exp\left[N\phi_{z,w,u}(\varepsilon,m)\right],
\end{equation}
where
\begin{equation}\label{eq:phizwu}
\fl \phi_{z,w,u}(\varepsilon,m)=z\varepsilon+wm+u+\frac{w^2}{4(u-zdJ)}-\frac{1}{2}\int_{[0,2\pi)^d}\frac{\dd^d\varphi}{(2\pi)^d}\,\ln\bigg(u-zJ\sum_{j=1}^d\cos\varphi_j\bigg).
\end{equation}
From this expression, the microcanonical entropy in the thermodynamic limit,\footnote{Note that here and in the following Boltzmann's constant is set to unity.}
\begin{equation}\label{eq:semtdl}
s(\varepsilon,m)=\lim_{N\to\infty}\frac{1}{N}\ln\Omega_N(\varepsilon,m),
\end{equation}
can be computed by the method of steepest descent (see section 6.6 of reference \cite{BenOrs}). Behringer proposes to do that by studying the saddle points of $\phi_{z,w,u}$ in the complex $(z,w,u)$-space (see equations (28)--(30) of reference \cite{Behringer:05}), and for his purposes such an analysis turns out to be sufficient. To obtain the entropy $s$ in the entire range of accessible values in the $(\varepsilon,m)$-plane, however, a saddle point analysis is not satisfactory. The reason for that, as we will see in the following, is that branch cuts in the complex $(z,u)$-space due to the logarithm in \eref{eq:phizwu} come into play.

We start by observing that the expression for the density of states $\Omega_N$ can be simplified by performing the $w$- and the $u$-integration in \eref{eq:Omega_3int}, yielding
\begin{eqnarray}
\fl \Omega_N(\varepsilon,m)\sim& \frac{N^{(N-5)/2}}{\sqrt{N\pi}\,\Gamma[(N-3)/2]}\int_{a-\ii\infty}^{a+\ii\infty}\frac{\dd z}{2\pi}
\sqrt{\frac{1-zdJ}{\left[1-m^2+z(\varepsilon+m^2dJ)\right]^5}}\nonumber\\
\fl &\times\exp\bigg[-\frac{N}{2\pi^d}\int_{[0,\pi)^d}\dd^d\varphi\,\ln\bigg(\frac{1-zJ\sum_{j=1}^d\cos\varphi_j}{1-m^2+z(\varepsilon+m^2dJ)}\bigg)\bigg]\label{eq:Omega_1int}
\end{eqnarray}
(the derivation of this result is given in \ref{appendixA}). Due to the logarithm in \eref{eq:Omega_1int}, the integrand of the $z$-integration has two branch cuts on the real line, one extending from $-\infty$ to $-1/(dJ)$, the other from $+1/(dJ)$ to $+\infty$. Apart from these branch cuts, the integrand is holomorphic, and the contour of integration can be deformed freely, as long as it does not cross the cuts. For an asymptotic evaluation of the integral in the large-$N$ limit by means of the method of steepest descent, the path of integration in the complex plane is deformed such that its imaginary part becomes zero. In this limit, the value of the integral is given by the integrand of the $z$-integration in \eref{eq:Omega_1int} evaluated at the maximum along that path. This maximum is located in the interval
\begin{equation}
{\mathcal I}=[-1/(dJ),+1/(dJ)]
\end{equation}
of the real $z$-axis, and---depending on the values of $\varepsilon$ and $m$---it may be either a saddle point of the exponent in \eref{eq:Omega_1int}, or one of the boundary points $\pm1/(dJ)$.

\begin{figure}\center
\psfrag{Rez}{\scriptsize $\Re(z)$}
\psfrag{Imz}{\scriptsize $\Im(z)$}
\psfrag{-0.1}{\scriptsize $\!-0.1$}
\psfrag{0.1}{\scriptsize $0.1$}
\psfrag{0.2}{\scriptsize $0.2$}
\psfrag{-0.2}{\scriptsize $\!-0.2$}
\psfrag{0.4}{}
\psfrag{-0.4}{}
\psfrag{0.5}{\scriptsize $0.5$}
\psfrag{0.9}{\scriptsize $0.9$}
\psfrag{1.3}{\scriptsize $1.3$}
\psfrag{1.7}{\scriptsize $1.7$}
\psfrag{-}{\scriptsize $\!\!-$}
\includegraphics[width=8cm]{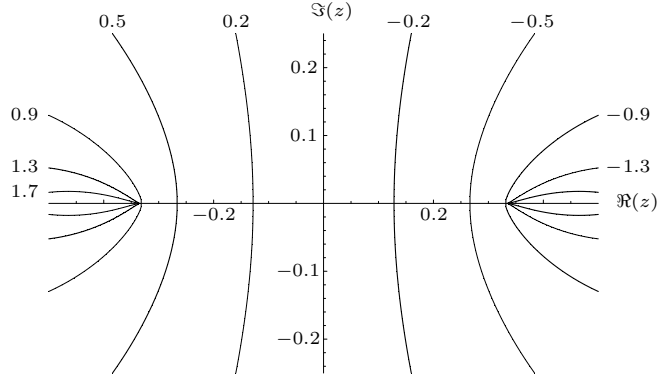}
\caption{\label{fig:paths}
Paths of vanishing imaginary part of the exponent of \eref{eq:Omega_1int} in the complex $z$-plane. The paths shown here are for the spherical model on a three-dimensional lattice with coupling constant $J=1$ and magnetization $m=0$, and for various values of the energy $\varepsilon$ as indicated in the plot. One can observe in this figure that, for intermediate values of $\varepsilon$, the paths cross the real axis in the interior of the interval $[-1/3,+1/3]$. For larger or smaller energies, the paths ``get stuck'' at the branch cuts at $-1/3$ and $+1/3$, respectively.
}
\end{figure}
\begin{figure}\center
\psfrag{e}{\small $\varepsilon$}
\psfrag{m}{\small $m$}
\psfrag{1}{\scriptsize $1$}
\psfrag{2}{\scriptsize $2$}
\psfrag{3}{\scriptsize $3$}
\psfrag{0.5}{\scriptsize $0.5$}
\psfrag{-}{\scriptsize $\!-$}
\psfrag{I}{\small I}
\psfrag{II}{\small II}
\psfrag{III}{\small III}
\vspace{-3cm}
\includegraphics[width=10cm]{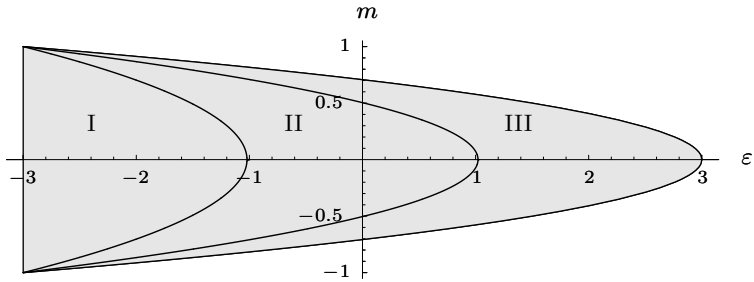}
\vspace{-3cm}
\caption{\label{fig:phasediagram}
Microcanonical phase diagram of the spherical model on a three-dimensional cubic lattice. The microcanonical entropy $s$ is defined on the grey shaded region in the $(\varepsilon,m)$-plane. Within each of the regions I, II, and III, the entropy is analytic, but not so on the boundaries separating the regions.
}
\end{figure}

%
For lattice dimensions $d=1$ and $d=2$, the maximum of the integrand along the path of vanishing imaginary part is always located at a saddle point in the interior of ${\mathcal I}$. As we will see from the following discussion, such a behaviour corresponds to the absence of a phase transition. For $d\geqslant3$, however, the situation is different: In \fref{fig:paths}, the behaviour of the paths of vanishing imaginary part upon variation of $\varepsilon$ and $m$ is illustrated for the spherical model on a lattice of dimension $d=3$ with coupling constant $J=1$, but the behaviour is similar for larger values of $d$. We can identify three different regions in the $(\varepsilon,m)$-plane which are sketched in \fref{fig:phasediagram}.
\begin{description}
\item[Region I:] For relatively small energies $\varepsilon$, the maximum of the integrand of the $z$-integration is located at $+1/(dJ)$, and the microcanonical entropy in the thermodynamic limit is given by
\begin{equation}\label{eq:sI}
\fl s^{\mathrm{I}}(\varepsilon,m)=\frac{1}{2}\ln\Big[2\ee\Big(1+\frac{\varepsilon}{dJ}\Big)\Big]-\frac{1}{2\pi^d}\int_{[0,\pi)^d}\dd^d\varphi\,\ln\bigg(1-\frac{1}{d}\sum_{j=1}^d\cos\varphi_j\bigg).
\end{equation}
Physically, region I corresponds to the {\em coexistence region}\/ in which domains of positive magnetization and domains of negative magnetization coexist. The independence of $s^{\mathrm{I}}$ on the magnetization $m$ in equation \eref{eq:sI} is a hallmark of this behaviour. This $m$-independence can be seen as a consequence of a remarkable interplay of numerator and denominator of the argument of the logarithm in \eref{eq:Omega_1int}: The numerator determines the locations of the branch cuts $(-\infty,-1/(dJ)]$ and $[+1/(dJ),+\infty)$ in the complex $z$-plane, and precisely at the endpoint $+1/(dJ)$ of one of these cuts the denominator, and therefore the entire exponent in \eref{eq:Omega_1int}, becomes independent of the magnetization $m$.
\item[Region II:] For intermediate energies, the integrand in \eref{eq:Omega_1int} has a saddle point $z_0$ in the interior of the interval ${\mathcal I}$, determined by the equation
\begin{equation}\label{eq:saddle}
\fl 0=\varepsilon+m^2dJ+\left[1-m^2+z_0(\varepsilon+m^2dJ)\right]\int_{[0,\pi)^d}\frac{\dd^d\varphi}{\pi^d}\,\frac{J\sum_{j=1}^d\cos\varphi_j}{1-z_0J\sum_{j=1}^d\cos\varphi_j}.
\end{equation}
The microcanonical entropy in the thermodynamic limit is then given by
\begin{eqnarray}
s^{\mathrm{II}}(\varepsilon,m)=&\frac{1}{2}\ln\Big[2\ee\Big(1-m^2+z_0(\varepsilon+m^2dJ)\Big)\Big]\nonumber\\
&-\frac{1}{2\pi^d}\int_{[0,\pi)^d}\dd^d\varphi\,\ln\bigg(1-z_0J\sum_{j=1}^d\cos\varphi_j\bigg).\label{eq:sII}
\end{eqnarray}
\item[Region III:]  For larger energies $\varepsilon$, the maximum of the integrand of the $z$-in\-te\-gra\-tion in \eref{eq:Omega_1int} is located at $-1/(dJ)$, and the microcanonical entropy in the thermodynamic limit is given by equation \eref{eq:sII} with $z_0=-1/(dJ)$.
\end{description}
The boundaries between the different regions in the $(\varepsilon,m)$-plane are given by the values of $(\varepsilon,m)$ for which the saddle point of the integrand in \eref{eq:Omega_1int} approaches one of the boundary points $z_0=\pm1/(dJ)$. These points are obtained by solving equation \eref{eq:saddle} with $z_0=\pm1/(dJ)$, yielding for the two boundary curves the parabolas
\begin{equation}\label{eq:epm}
\varepsilon_\pm(m)=-dJ\frac{m^2+a_d\left[m^2\pm(1-m^2)\right]}{1+a_d}
\end{equation}
with
\begin{equation}\label{eq:ad}
a_d=\int_{[0,\pi)^d}\frac{\dd^d\varphi}{\pi^d}\,\frac{\sum_{j=1}^d\cos\varphi_j}{d-\sum_{j=1}^d\cos\varphi_j}.
\end{equation}

\begin{figure}\center
\psfrag{0.5}{\small $m$}
\psfrag{e}{\small $\varepsilon$}
\psfrag{-3}{\scriptsize $\!-3$}
\psfrag{-2}{\scriptsize $\!-2$}
\psfrag{-1}{\scriptsize $\!-1$}
\psfrag{0}{\scriptsize $0$}
\psfrag{1}{\scriptsize $1$}
\psfrag{2}{\scriptsize $2$}
\psfrag{3}{\scriptsize $3$}
\includegraphics[width=8.6cm]{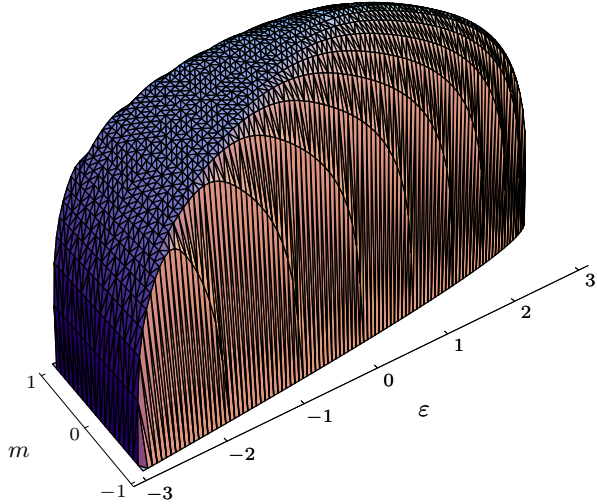}
\caption{\label{fig:entropy}
Graph of the microcanonical entropy $s(\varepsilon,m)$ of the spherical model with coupling strength $J=1$ on a three-dimensional cubic lattice in the thermodynamic limit. The constant (in the $m$-direction) part on the left hand side of the graph corresponds to region I where phase coexistence occurs.
}
\end{figure}

Since the microcanonical entropy $s(\varepsilon,m)$ in the thermodynamic limit is defined piecewise in the regions I, II, III, one expects this function to be nonanalytic (i.\,e.\ not infinitely-many times differentiable) at the borders of the regions. As a consequence, the plot in \fref{fig:phasediagram} can be interpreted as a microcanonical phase diagram, showing at which values of the control parameters $\varepsilon$ and $m$ thermodynamic singularities are encountered. From an overall plot of the graph of $s(\varepsilon,m)$ as shown in \fref{fig:entropy} this nonanalytic behaviour is not immediately visible and we will consider derivatives of $s$ to illustrate these features in the next section.

In summary, we have observed that the asymptotic evaluation of the integral governing the density of states yields different behaviours for the regions I, II, and III, corresponding to the different phases of the spherical model. Only for region II is a saddle point analysis of the integral in \eref{eq:Omega_1int} applicable. Upon reaching the boundaries of region II, the saddle point on the real $z$-axis gets stuck at the branch cut of the exponent in equation \eref{eq:Omega_1int}, leading to nonanalytic behaviour of the microcanonical entropy $s(\varepsilon,m)$. Remarkably, at the boundary $\varepsilon_+(m)$ between regions I and II, the saddle point gets stuck at a point precisely such that the $m$ dependence of the exponent in \eref{eq:Omega_1int} cancels, leading to an affine (in the $m$-direction) region of the entropy.


\section{Caloric curves and specific heat at fixed magnetization}
\label{sec:response}

\begin{figure}\center
\psfrag{T}{\small $T$}
\psfrag{b}{\small $\beta$}
\psfrag{e}{\small $\varepsilon$}
\psfrag{-10}{\scriptsize $-10$}
\psfrag{-5}{\scriptsize $-5$}
\psfrag{5}{\scriptsize $5$}
\psfrag{10}{\scriptsize $10$}
\psfrag{1}{\scriptsize $1$}
\psfrag{2}{\scriptsize $2$}
\psfrag{0.5}{\scriptsize $0.5$}
\psfrag{1.5}{\scriptsize $1.5$}
\psfrag{2.9}{\scriptsize $2.9$}
\psfrag{2.8}{\scriptsize $2.8$}
\psfrag{2.7}{\scriptsize $2.7$}
\psfrag{2.6}{\scriptsize $2.6$}
\psfrag{2.4}{\scriptsize $2.4$}
\psfrag{2.2}{\scriptsize $2.2$}
\psfrag{-}{\scriptsize $\!-$}
\includegraphics[width=6cm]{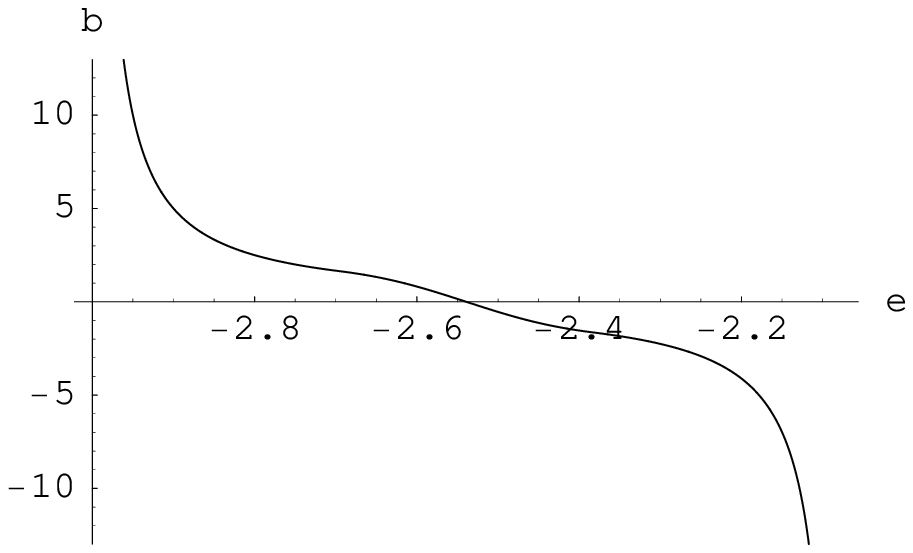}
\hfill
\includegraphics[width=6cm]{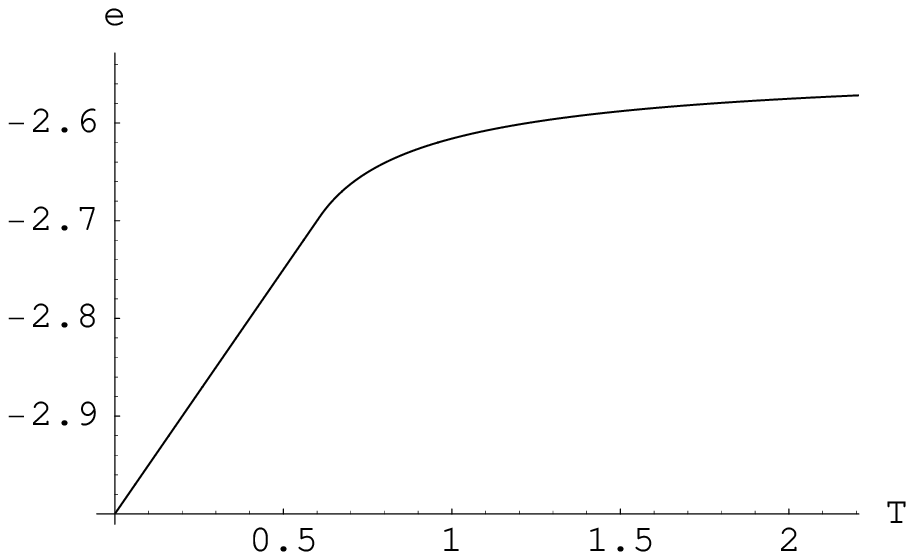}
\caption{\label{fig:caloric}
Microcanonical caloric curves of the spherical model on a three-di\-men\-sio\-nal cubic lattice with fixed magnetization $m_0=0.92$ in the thermodynamic limit. Left: Inverse temperature $\beta$ as a function of the energy $\varepsilon$ as defined in \eref{eq:caloric}. Right: A more familiar representation of the caloric curve, showing $\varepsilon$ as a function of the temperature $T=1/\beta$, restricted to positive values of $T$.
}
\end{figure}

\begin{figure}\center
\psfrag{c}{\small $c$}
\psfrag{s}{\small $s_{m_0}''$}
\psfrag{''}{}
\psfrag{e}{\small $\varepsilon$}
\psfrag{18}{\scriptsize $18$}
\psfrag{16}{\scriptsize $16$}
\psfrag{14}{\scriptsize $14$}
\psfrag{12}{\scriptsize $12$}
\psfrag{10}{\scriptsize $10$}
\psfrag{6}{\scriptsize $6$}
\psfrag{0.5}{\scriptsize $0.5$}
\psfrag{0.4}{\scriptsize $0.4$}
\psfrag{0.3}{\scriptsize $0.3$}
\psfrag{0.2}{\scriptsize $0.2$}
\psfrag{0.1}{\scriptsize $0.1$}
\psfrag{2.8}{\scriptsize $2.8$}
\psfrag{2.6}{\scriptsize $2.6$}
\psfrag{2.4}{\scriptsize $2.4$}
\psfrag{2.2}{\scriptsize $2.2$}
\psfrag{-}{\scriptsize $\!-$}
\includegraphics[width=6cm]{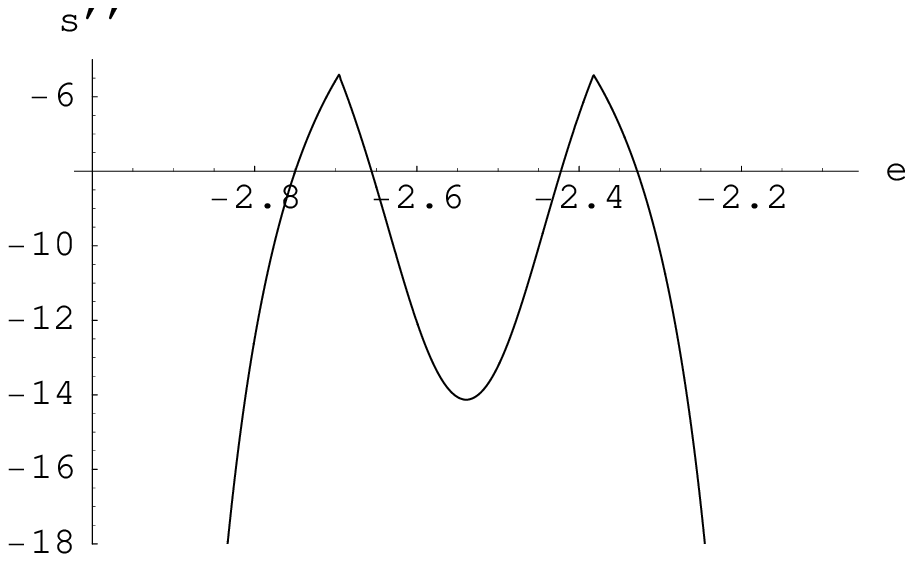}
\hfill
\includegraphics[width=6cm]{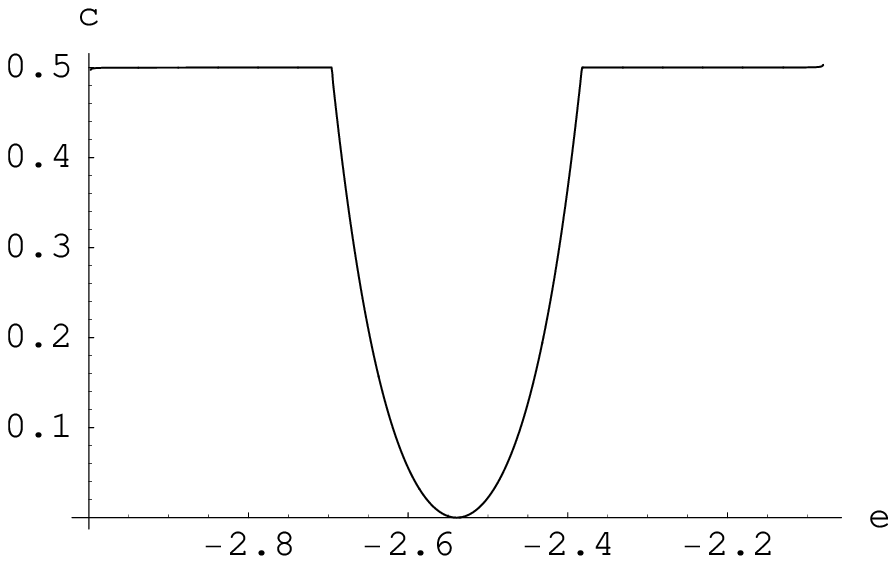}
\caption{\label{fig:specificheat}
Left: Second derivative $s''_{m_0}(\varepsilon)$ of the entropy at fixed magnetization $m_0=0.92$ for the spherical model on a three-dimensional cubic lattice in the thermodynamic limit. Right: Microcanonical specific heat $c(\varepsilon)$ as defined in \eref{eq:c} for the same model and parameter values.
}
\end{figure}

Caloric curves or response functions like the specific heat are typical quantities which are measured in experiments, and kinks, discontinuities, or singularities in these quantities are hallmarks of a phase transitions. From the microcanonical entropy, these quantities are obtained as first or second derivatives (or functions thereof). To illustrate the occurrence of nonanalyticities in the microcanonical entropy $s(\varepsilon,m)$, we consider the spherical model at fixed magnetization $m_0$. The corresponding microcanonical entropy function is
\begin{equation}
s_{m_0}(\varepsilon)=s(\varepsilon,m_0),
\end{equation}
and the microcanonical caloric curve is given as 
\begin{equation}\label{eq:caloric}
\beta(\varepsilon)=\frac{\dd s_{m_0}(\varepsilon)}{\dd \varepsilon},
\end{equation}
where $\beta=1/T$ is the inverse temperature. For the magnetization $m_0=0.92$ the caloric curve is plotted in \fref{fig:caloric}.
The nonanalyticity of the microcanonical entropy $s$ is clearly visible only in the second derivative
\begin{equation}\label{eq:d2sde2}
s_{m_0}''(\varepsilon)=\frac{\dd^2 s_{m_0}(\varepsilon)}{\dd \varepsilon^2},
\end{equation}
or in the microcanonical specific heat
\begin{equation}\label{eq:c}
c(\varepsilon)=-\frac{[s_{m_0}'(\varepsilon)]^2}{s_{m_0}''(\varepsilon)},
\end{equation}
both plotted in \fref{fig:specificheat}.
Qualitatively, graphs for other values of the fixed magnetization $m_0$ look similar.

The observed nonanalyticities of $s(\varepsilon,m)$ can also be interpreted as phase transitions in the spherical model with fixed magnetization. For any value of the fixed magnetization $m_0$, two continuous phase transitions occur at the values $\varepsilon_\pm(m_0)$ as given in equation \eref{eq:epm}. The order of the phase transition, i.\,e.\ the fact that it is a continuous one, could have been deduced also without computing the exact result for $s(\varepsilon,m)$ from geometric arguments similar to those applied to the Ising model in reference \cite{Kastner:02}.

\section{Legendre transform and canonical free energy}
\label{sec:legendre}
The microcanonical entropy that we have computed in \sref{sec:micent} forms the starting point of an analysis of the spherical model in the microcanonical ensemble. An analogous role is played by the canonical Gibbs free energy 
\begin{equation}
g_N(\beta,h)=-\frac{1}{N\beta}\int_{\RR^N}\dd\sigma\,\Dirac\bigg(N-\sum_{i=1}^N\sigma_i^2\bigg)\ee^{-\beta H_N(\sigma)}
\end{equation}
for calculations in the canonical ensemble. In the thermodynamic limit, the corresponding infinite-system quantity
\begin{equation}
g(\beta,h)=\lim_{N\to\infty}g_N(\beta,h)
\end{equation}
is related to the microcanonical entropy $s(\varepsilon,m)$ by means of a Legendre transform,
\begin{equation}
-\beta g(\beta,h)=s(\bar{\varepsilon}(\beta,h),\bar{m}(\beta,h))-\beta\bar{\varepsilon}(\beta,h)+\beta h\bar{m}(\beta,h),
\end{equation}
where $\bar{\varepsilon}(\beta,h)$ and $\bar{m}(\beta,h)$ are the solutions of the equations
\begin{equation}\label{eq:ebetamh}
\frac{\partial}{\partial\bar{\varepsilon}}s(\bar{\varepsilon},\bar{m})-\beta=0,\qquad \frac{\partial}{\partial\bar{m}}s(\bar{\varepsilon},\bar{m})+\beta h=0.
\end{equation}
In this way we can derive the canonical result from the microcanonical one. While, in the thermodynamic limit, the canonical free energy is always given as the Legendre transform of the microcanonical entropy, the inverse is not always true: Equivalence of ensembles holds only for concave entropy functions \cite{ElHaTur:00,TouElTur:04}, and only in this case can the entropy be recovered from the free energy by means of a Legendre transform. 

\begin{figure}\center
\includegraphics[width=7cm]{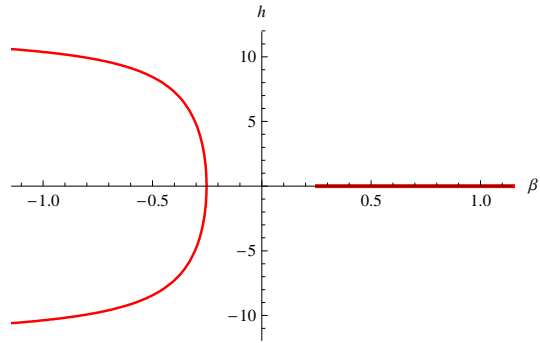}
\caption{\label{fig:canPD}
Canonical phase diagram of the spherical model. The plotted lines in the $(\beta,h)$-plane mark the parameter values at which the canonical free energy $f(\beta,h)$ is nonanalytic and phase transitions take place. The right line corresponds to the transition line $\varepsilon_+(m)$ in the microcanonical case at which a transition from paramagnetic to ferromagnetic behaviour takes place. The left line corresponds to the transition line $\varepsilon_-(m)$ in the microcanonical case at which a transition from antiferromagnetic to paramagnetic behaviour occurs.
}
\end{figure}

The equations in \eref{eq:ebetamh} provide a way to translate the microcanonical control parameters $(\varepsilon,m)$ into the canonical control parameters $(\beta,h)$. We can use these relations to translate the microcanonical transition lines \eref{eq:epm} in the $(\varepsilon,m)$-plane into the canonical transition lines in the $(\beta,h)$-plane. From the microcanonical transition line $\varepsilon_+(m)$ in equation \eref{eq:ebetamh} we obtain the canonical transition line $h_+(\beta)=0$ for $\beta\in[(1+a_d)/2dJ,+\infty)$, with $a_d$ as defined in equation \eref{eq:ad}. From the transition line $\varepsilon_-(m)$ we obtain the canonical counterpart
\begin{equation}
\beta_-(h)=\frac{8dJ(1+a_d)}{h^2-(4dJ)^2}
\end{equation}
for $h\in\RR$. Details of the calculation are given in \ref{app:canPD}, and the resulting canonical phase diagram is plotted in \fref{fig:canPD}.

The two transition lines in the $(\beta,h)$-plane display the typical behaviour of a ferromagnetic transition (for positive $\beta$) and of an antiferromagnetic transition (for negative $\beta$), and this observation deserves some remarks. For physical reasons, the canonical ensemble is often restricted to positive temperatures, corresponding to positive values of the inverse temperature $\beta$. First, this appears to be appropriate for modelling a system in a thermal bath, but there is a second reason for this restriction: For many physical systems, in particular those with a standard kinetic energy term which is a quadratic form in the momenta, no canonical equilibrium states corresponding to negative temperatures exist. This is different for the spherical model, as well as for many other spin models for which the energy per degree of freedom is bounded above.

Imagining the physical situation of a macroscopically large spherical model coupled to a thermal reservoir with positive inverse temperature $\beta$, only the right half-plane in \fref{fig:canPD} can be explored and the ferromagnetic-to-paramagnetic transition can be probed. The left half-plane, corresponding to negative temperatures, might still be accessed by means of a trick: Since, in the canonical ensemble, only the product $\beta H_N$ of $\beta$ times the Hamiltonian function $H_N$ enters the calculation of thermodynamic quantities, the thermodynamics of a system with Hamiltonian $H_N$ at negative temperatures is identical to that of a system with Hamiltonian $-H_N$ at positive temperatures. For the spherical model, this simply corresponds to an inversion of the sign of the coupling constant $J$ in the Hamiltonian function \eref{eq:H_sph}, giving rise to a spin system with antiferromagnetic coupling $J<0$. This reasoning explains the appearance of a transition line of antiferromagnetic type in the left half-plane of \fref{fig:canPD}.

It also follows from this discussion that equivalence of the microcanonical and the canonical results for the spherical model requires the canonical ensemble to be considered for positive as well as negative temperatures: Only then it is possible to derive from the canonical free energy, by means of a Legendre transform, the microcanonical entropy $s(\varepsilon,m)$ on its entire domain, implying equivalence of the microcanonical and the canonical ensemble on the thermodynamic level \cite{ElHaTur:00,TouElTur:04}. Note that this is a common situation for spin models, occurring for the Ising model, the Potts model, and many others.

\section{Discussion and conclusions}
\label{sec:conclusions}

We have presented a calculation of the microcanonical entropy $s(\varepsilon,m)$ of the spherical model with nearest-neighbour interaction, valid for all accessible values of the energy $\varepsilon$ and the magnetization $m$ per spin. The entropy function is found to be concave (albeit not strictly concave) and, as a consequence, equivalence of the microcanonical and the canonical ensemble holds. Since, in addition to the nearest-neighbour interactions, the degrees of freedom of the spherical model are subject to an---effectively long-range---constraint, equivalence of ensembles could not have been taken for granted from the outset. We can interpret our finding of a concave entropy, concluding that the long-range constraint is not sufficient to change the overall characteristics of the model into that of a long-range interacting one. Only when the interactions between the spins are made properly long-range (like in the mean-field spherical model discussed in \cite{Kastner:06,KaSchne:06}) would we expect the entropy $s(\varepsilon,m)$ to develop a nonconcave part and nonequivalence of ensembles to occur.

When computing the microcanonical entropy of the spherical model by means of the method of steepest descent in \sref{sec:micent}, we have observed different regions, depending on the values of the parameters $\varepsilon$ and $m$. In region II, the main contribution to the relevant asymptotic integral comes from a proper saddle point on the real axis of the integration variable $z$. In regions I and III, however, the saddle point gets stuck at a branch cut in the complex $z$-plane. It is this ``getting-stuck'' which leads to nonanalytic behaviour of $s$, giving rise to a phase transition in the spherical model.

Note that a similar mechanism of generating nonanalytic behaviour from the interplay of a saddle point and a branch cut has been observed previously. In the very different context of nonequilibrium stationary states described by a Langevin equation, Farago \cite{Farago:02} calculates the probability distribution function of the injection of energy into the system. His calculation also makes use of the method of steepest descent and, depending on some parameter values, a saddle point can approach a branch cut in the complex plane (Figure 3 of \cite{Farago:02}). Once the saddle points gets stuck there, the probability distribution function develops a nonanalyticity, similar to what we observed for the microcanonical entropy of the spherical model. Farago also remarks that this nonanalyticity can be interpreted as a phase transition; for details see the discussion in section 2.3 of reference \cite{Farago:02}.

The lines in the $(\varepsilon,m)$-plane at which the microcanonical entropy is nonanalytic, plotted in \fref{fig:phasediagram}, can be interpreted as a microcanonical phase diagram. Compared to its canonical counterpart shown in \fref{fig:canPD}, the microcanonical phase diagram looks remarkably different, and also its physical implications are unfamiliar when compared to the canonical situation: Varying, for example, the energy $\varepsilon$ while keeping fixed the magnetization at any value of $m$, one typically crosses {\em two}\/ transition lines in the phase diagram, therefore observing two phase transitions, signaled by kinks in the specific heat. Similarly, four transition lines are crossed upon variation of the magnetization while keeping the energy $\varepsilon$ fixed at any value $-d<\varepsilon<\varepsilon_+(0)$, while two transition lines are crossed for energies $\varepsilon_+(0)<\varepsilon<\varepsilon_-(0)$. Qualitatively similar results can also obtained for the Ising model \cite{KaPlei:09}, and presumably for many other short-range spin models. For more details, especially on physical realizations of such models under microcanonical conditions, the reader is referred to reference \cite{KaPlei:09}.

\section*{Acknowledgments}
I would like to thank Oliver Schnetz for useful hints how to tackle the integral \eref{eq:Omega_3int}, and Hugo Touchette for helpful discussions and for pointing out reference \cite{Farago:02}. 

\appendix

\section{Derivation of equation \eref{eq:Omega_1int}}
\label{appendixA}

Starting from equation \eref{eq:Omega_3int} with exponent \eref{eq:phizwu}, the $w$-integration is a Gaussian integral which is straightforward to perform:
\begin{eqnarray}
\fl \Omega_N(\varepsilon,m)\sim\int_{a-\ii\infty}^{a+\ii\infty}\frac{\dd z}{2\pi} \int_{c-\ii\infty}^{c+\ii\infty}\frac{\dd u}{2\pi} \int_{b-\ii\infty}^{b+\ii\infty}\frac{\dd w}{2\pi}\nonumber\\
\fl \times\exp\bigg\{N\bigg[\frac{w^2}{4(u-zdJ)}+wm+z\varepsilon+u-\frac{1}{2}\int_{[0,2\pi)^d}\frac{\dd^d\varphi}{(2\pi)^d}\,\ln\bigg(u-zJ\sum_{j=1}^d\cos\varphi_j\bigg)\bigg] \bigg\}\nonumber\\
\fl = \int_{a-\ii\infty}^{a+\ii\infty}\frac{\dd z}{2\pi} \int_{c-\ii\infty}^{c+\ii\infty}\frac{\dd u}{2\pi} \exp\!\bigg[N\bigg(f(u,z)-\frac{m^2}{4a}\bigg)\bigg]\int_{b-\ii\infty}^{b+\ii\infty}\frac{\dd w}{2\pi} \exp\!\bigg[N\bigg(w\sqrt{a}+\frac{m}{2\sqrt{a}}\bigg)^2\bigg]\nonumber\\
\fl = \frac{\ii}{2\sqrt{N\pi}}\int_{a-\ii\infty}^{a+\ii\infty}\frac{\dd z}{2\pi} \int_{c-\ii\infty}^{c+\ii\infty}\frac{\dd u}{2\pi} \frac{1}{\sqrt{a(u,z)}}\exp\left[N\left(f(u,z)-\frac{m^2}{4a(u,z)}\right)\right]\label{eq:Gauss}
\end{eqnarray}
with
\begin{equation}
a(u,z)=\frac{1}{4(u-zdJ)}
\end{equation}
and
\begin{equation}
f(u,z)=u+z\varepsilon-\frac{1}{2}\int_{[0,2\pi)^d}\frac{\dd^d\varphi}{(2\pi)^d}\,\ln\bigg(u-zJ\sum_{j=1}^d\cos\varphi_j\bigg).
\end{equation}
The Gaussian integral in \eref{eq:Gauss} has been solved assuming that $\Re(u)>\Re(zdJ)$ along the contour of integration. To solve the $u$-integral in \eref{eq:Gauss}, we perform the substitution of variables
\begin{equation}
z\rightarrow xu,
\end{equation}
yielding
\begin{eqnarray}
\fl \Omega_N(\varepsilon,m)\sim\frac{\ii}{\sqrt{N\pi}}\int\frac{\dd x}{2\pi} \sqrt{1-xdJ} \exp\bigg\{-\frac{N}{2}\int_{[0,2\pi)^d}\frac{\dd^d\varphi}{(2\pi)^d}\,\ln\bigg(1-xJ\sum_{j=1}^d\cos\varphi_j\bigg)\bigg\}\nonumber\\
\times \int\frac{\dd u}{2\pi} u^{(3-N)/2}\exp\left\{Nu\left[1+x\varepsilon-m^2(1-xdJ)\right]\right\}.\label{eq:intermediate}
\end{eqnarray}
In a second step we substitute
\begin{equation}
Nu\left[1+x\varepsilon-m^2(1-xdJ)\right]\rightarrow y
\end{equation}
in the second integral in \eref{eq:intermediate}, yielding
\begin{eqnarray}
\fl \Omega_N(\varepsilon,m)\sim\frac{\ii}{\sqrt{N\pi}}N^{(N-5)/2} \int\frac{\dd x}{2\pi} \sqrt{\frac{1-xdJ}{\left[1+x\varepsilon-m^2(1-xdJ)\right]^{N-5}}}\nonumber\\
\fl \quad\times\exp\bigg\{-\frac{N}{2}\int_{[0,2\pi)^d}\frac{\dd^d\varphi}{(2\pi)^d}\,\ln\bigg(1-xJ\sum_{j=1}^d\cos\varphi_j\bigg)\bigg\} g(N),\label{eq:almostfinal}
\end{eqnarray}
where
\begin{equation}\label{eq:gN}
g(N)=\int\frac{\dd y}{2\pi}y^{(3-N)/2}\ee^y.
\end{equation}
The $y$-integration is along some path extending from $\alpha-\ii\infty$ to $\alpha+\ii\infty$ for $\alpha>0$.

For $(N-3)/2\in\NN$, the integrand in \eref{eq:gN} has a single pole at the origin, and by means of the residue theorem we obtain
\begin{equation}
g(N)=\frac{\ii}{[(N-5)/2]!}=\frac{\ii}{\Gamma[(N-3)/2]}.
\end{equation}

For $(N-4)/2\in\NN$, the integrand in \eref{eq:gN} has a branch cut along the negative real axis. The contour of integration is deformed such that---for some small $\delta>0$---it goes from $\alpha-\ii\infty$ to $-\infty-\ii\delta$, follows slightly below the branch cut and back to $-\infty+\ii\delta$ slightly above the cut, and then to $\alpha+\ii\infty$. Only the contours along the branch cut contribute to the integral, yielding integral representations of the $\Gamma$-function. In summary, we obtain again
\begin{equation}
g(N)=\frac{2\ii}{2\pi}\Gamma[(5-N)/2]=\frac{\ii}{\Gamma[(N-3)/2]}
\end{equation}

Inserting these results into \eref{eq:almostfinal} and rearrangement of the terms leads to equation \eref{eq:Omega_1int}.

\section{Derivation of the canonical phase diagram}
\label{app:canPD}

In this appendix we derive the canonical phase diagram as shown in \fref{fig:canPD} from the microcanonical one in \fref{fig:phasediagram}. For this purpose, we translate the microcanonical transition lines $\varepsilon_\pm(m)$ from equation \eref{eq:epm} into the corresponding values of $\beta$ and $h$ by means of the equations in \eref{eq:ebetamh}.

\paragraph{Transition line $\varepsilon_+(m)$:} This transition line separates the regions I and II as defined in \sref{sec:micent}. On the boundary we can therefore use either $s^{\mathrm{I}}$ from equation \eref{eq:sI} or $s^{\mathrm{II}}$ from equation \eref{eq:sII}, and for simplicity we choose the former. Inserting $s^{\mathrm{I}}$ into the equations in \eref{eq:ebetamh} we obtain
\begin{eqnarray}
\frac{\partial}{\partial m}s^{\mathrm{I}}(\varepsilon,m)&=&0=-\beta h,\\
\frac{\partial}{\partial\varepsilon}s^{\mathrm{I}}(\varepsilon,m)&=&\frac{1}{2(dJ+\varepsilon)}=\beta(\varepsilon).\label{eq:betae}
\end{eqnarray}
As a consequence, the canonical transition line is located at $h=0$. Evaluating equation \eref{eq:betae} at the microcanonical transition line $\varepsilon_+(m)$, we notice that $\varepsilon_+$ can take on values ranging from $-dJ$ to $\varepsilon_+(0)=-dJa_d/(1+a_d)$, which implies that
\begin{equation}
\beta\in[\beta(\varepsilon_+(0)),\beta(-dJ))=\Bigl[\frac{1+a_d}{2dJ},+\infty\Bigr)
\end{equation}
with $a_d$ as defined in equation \eref{eq:ad}.

\paragraph{Transition line $\varepsilon_-(m)$:} This transition line separates the regions II and III. We choose to insert $s^{\mathrm{III}}$---given by equation \eref{eq:sII} with $z_0=-1/(dJ)$---into the equations in \eref{eq:ebetamh}, yielding
\begin{eqnarray}
\frac{\partial}{\partial m}s^{\mathrm{III}}(\varepsilon,m)&=&-\frac{4mdJ}{2dJ(1-2m^2)-2\varepsilon}=-\beta h(\varepsilon,m),\label{eq:smIII}\\
\frac{\partial}{\partial\varepsilon}s^{\mathrm{III}}(\varepsilon,m)&=&-\frac{1}{2dJ(1-2m^2)-2\varepsilon}=\beta(\varepsilon,m).\label{eq:seIII}
\end{eqnarray}
Dividing \eref{eq:smIII} by \eref{eq:seIII} we obtain
\begin{equation}\label{eq:hofm}
h=-4dJm.
\end{equation}
Now we evaluate \eref{eq:seIII} at $\varepsilon_-(m)$,
\begin{equation}\label{eq:betaminus}
\beta_-(m):=\beta(\varepsilon_-(m),m)=\frac{1+a_d}{2dJ(m^2-1)},
\end{equation}
and by inserting expression \eref{eq:hofm} into \eref{eq:betaminus} we obtain the canonical transition line
\begin{equation}
\beta_-(h)=\frac{8dJ(1+a_d)}{h^2-(4dJ)^2}.
\end{equation}

\vspace{4mm}
\bibliographystyle{unsrt}
\bibliography{SphericalNN}

\end{document}